\documentclass[12pt]{article}

\begin{document}
\title{Parity violation through color superconductivity}

\author{R.D.\ Pisarski$^1$ and D.H.\ Rischke$^2$ \\ ~~ \\
$^1${\small Physics Dept., 
Brookhaven National Laboratory,} \\
{\small Upton, NY 11973--5000, U.S.A.} \\
{\small \it email: pisarski@bnl.gov} \\
$^2${\small RIKEN-BNL Research
Center, Brookhaven National Laboratory,} \\ 
{\small Upton, NY 11973--5000, U.S.A.} \\
{\small \it email: rischke@bnl.gov} }


\maketitle
\begin{abstract}
We give a pedagogical discussion of how color superconductivity 
can produce parity violation in cold quark matter at very high densities.
\end{abstract}

In this note, we give a pedagogical discussion of how, for massless quarks
at very high densities, the formation of a spin-zero color superconducting
condensate spontaneously breaks both the axial $U(1)$ symmetry and parity
\cite{pisris:pr}.
This observation is implicit in the seminal work of Bailin and Love, 
is noted by Alford, Rajagopal, and Wilczek, and is
explicitly discussed by Evans, Hsu, and Schwetz \cite{pisris:bl}.

For simplicity, consider 
two degenerate flavors of quarks, and assume that a quark-quark condensate
forms in the color-antitriplet channel 
\cite{pisris:pr,pisris:bl,pisris:2,pisris:br}.
For massless quarks, 
two of the four possible condensates with total spin $J=0$ are 
\cite{pisris:pr}
\begin{equation}
\langle \phi^a_1\rangle  = 
\epsilon^{abc} \, \epsilon_{fg}\, \langle \,{q^b_f}^T \, C \,
\gamma_5 \, q^c_g \, \rangle \,\,\,\, {\rm and} \,\,\,\,
\langle \phi^a_2 \rangle = 
\epsilon^{abc} \, \epsilon_{fg}\, \langle \,{q^b_f}^T \, C \,
{\bf 1} \, q^c_g \, \rangle \,\,,
\end{equation}
where $a,b,c=1,2,3$ are $SU(3)_c$ color indices, $f,g=1,2$ are
$SU(2)_f$ flavor indices, and $C$ is the charge conjugation matrix.
$\phi^a_{1,2}$ are antitriplets under $SU(3)_c$ gauge transformations and
singlets under $SU(2)_f$ rotations \cite{pisris:pr}. The condensate $\phi^a_1$
is even under parity, $J^P=0^+$, while $\phi^a_2$ is odd,
$J^P=0^-$. There are two other condensates \cite{pisris:pr}, but 
they do not change our qualitative arguments about parity violation,
and so we omit them.

In the limit where mass and instanton-induced terms can be neglected,
the effective Lagrangian for color superconductivity is
\begin{equation} \label{pisris:eq1}
{\cal L}_0 = \left| \partial_\mu \phi_1 \right|^2 + 
\left| \partial_\mu \phi_2 \right|^2 +
\lambda \left( \left| \phi_1 \right|^2 + \left| \phi_2 \right|^2
- |v|^2 \right)^2  \,\, ,
\end{equation}
where $|\phi|^2 \equiv \sum_a (\phi^a)^* \phi^a$.
When mass and instanton effects are neglected,
the Lagrangian is symmetric under axial $U(1)$ transformations, which
rotate $\phi^a_1$ and $\phi^a_2$ into each other. Therefore, there is
only one quartic coupling, $\lambda$. The Lagrangian (\ref{pisris:eq1})
generates nonzero
vacuum expectation values for the $\phi^a$'s, which can be written as
\begin{equation}
\langle \phi^a_1 \rangle = v^a\, \cos \theta \,\,\,\, , \,\,\,\,
\langle \phi^a_2 \rangle = v^a\, \sin \theta \,\,.
\end{equation}
Condensation picks out a given direction in color space for $v^a$, and
a given value for $\theta$. $v^a \neq 0$ breaks the $SU(3)_c$ color 
symmetry, which produces color
superconductivity. 
$\theta \neq 0$ breaks the axial $U(1)$ symmetry. Further,
whenever $\theta \neq 0$, there is
a nonzero $J^P=0^-$ condensate $\langle \phi^a_2 \rangle$;
this represents the spontaneous breaking of parity
(relative to the external vacuum).

This breaking of parity is actually familiar from the spontaneous breaking of
chiral symmetry. Consider two flavors of {\em massless\/}
quarks; the effective potential is $O(4)$-symmetric, involving the
$J^P=0^+$ $\sigma$- and $J^P=0^-$ $\pi$-meson fields. For massless quarks, it
is as likely for a parity-odd pion condensate to form as it is for a 
parity-even $\sigma$-meson condensate. This does not happen in nature,
because nonzero quark masses break chiral symmetry explicitly,
and thus favor a $0^+$ condensate.

Similarly, it is important to add to the effective Lagrangian 
(\ref{pisris:eq1}) terms which explicitly
break the axial $U(1)$ symmetry:
\begin{equation}
{\cal L}' = 
- c \left( \left| \phi_1 \right|^2 - \left| \phi_2 \right|^2 \right)
+ m^2 \left| \phi_2 \right|^2 \,\, .
\end{equation}
As shown by Berges and Rajagopal \cite{pisris:br}, 
the first term is due to instantons, with $c$ proportional to the 
instanton density. 
Instantons are attractive in the $J^P=0^+$ channel, and repulsive in 
the $J^P=0^-$ channel, so $c$ is positive. 

In the second term, each power of the current quark mass $m_q$ is accompanied
by one power of $\phi^a_2$. Since $\phi^a_2$ itself is not gauge invariant,
the simplest gauge-invariant term is $m_q^2 |\phi_2|^2$
\cite{pisris:pr}, so $m \sim m_q$.
Thus, the pseudo-Goldstone boson for the axial $U(1)$ symmetry is
extremely light, $m \sim 10$ MeV, taking $m_q$ to be 
the up or down quark mass and assuming the constant of proportionality between
$m$ and $m_q$ to be of order 1. 
This is in contrast to the explicit breaking of chiral symmetry,
where the corresponding term is linear in the quark mass. 
The pseudo-Goldstone bosons are the pions which are relatively heavy,
$m_\pi \simeq 140 \, {\rm MeV} \sim \sqrt{m_q}$.

Both instanton and mass terms act to favor 
the formation of the $0^+$ condensate $\phi_1$ 
over that of the $0^-$ condensate $\phi_2$. Consider, however, the
limit of very high densities. When the quark 
chemical potential $\mu \rightarrow \infty$, the instanton density and so 
$c$ vanish like $\sim \mu^{-29/3}$ (for two flavors). 
The real question is whether at some density the current
quark mass is negligible compared to the scale of the condensate. If this
happens, we reach an
``instanton-free'' region in which quarks are effectively massless, 
${\cal L}'$ can be neglected, and parity is spontaneously broken.

Because mass terms are always present, the true thermodynamic ground state is
always the parity-even $0^+$ condensate, i.e., $\theta=0$. There is, however,
a finite probability for the system to condense in a parity-odd
state, i.e., $\theta \neq 0$. The size and lifetime of this state
is set by the mass of the pseudo-Goldstone bosons. 
For chiral symmetry breaking, the characteristic
scale is $1/m_\pi \sim 1.4$ fm. This is 
small compared to the time and length scales of a heavy-ion collision, so
that parity-odd fluctuations average to zero. On the other
hand, the region in space-time over which a parity-odd 
color superconducting condensate forms
is large, $1/m \sim 20$ fm.
If the collision time is shorter than this time
scale, there is a finite probability that the system decays in
a parity-odd state. 
We therefore propose to trigger on phase-space regions where nuclear
matter is cool and dense, in order to observe
the formation of parity-odd color-superconducting
condensates on an event-by-event basis.
A possible global parity-odd observable was discussed in 
\cite{pisris:kpt}.

\end{document}